\pdfoutput=1

\documentclass[11pt,twoside,a4paper,cmspaper,final,collab]{cms-tdr}
\def\svnVersion{177020}\def\cmsCernNo{2012-285}\def\cmsCernDate{\today}\def\cmsMessage{Submitted to Physics Letters B}

\begin{document}\cmsNoteHeader{TOP-11-003}

\hyphenation{had-ron-i-za-tion}
\hyphenation{cal-or-i-me-ter}
\hyphenation{de-vices}

\RCS$Revision: 170996 $
\RCS$HeadURL: svn+ssh://svn.cern.ch/reps/tdr2/papers/TOP-11-003/trunk/TOP-11-003.tex $
\RCS$Id: TOP-11-003.tex 170996 2013-02-11 18:19:52Z alverson $
\providecommand{\lum}{\ensuremath{\,\text{(lum.)}}\xspace}
\newlength\cmsFigWidth
\ifthenelse{\boolean{cms@external}}{\setlength\cmsFigWidth{0.95\columnwidth}}{\setlength\cmsFigWidth{0.8\textwidth}}
\cmsNoteHeader{TOP-11-003}
\title{Measurement of the \ttbar production cross section in pp collisions at $\sqrt{s}=7$\TeV with lepton + jets final states}

\date{\today}

\abstract{A measurement of the \ttbar production cross
section  in pp collisions at $\sqrt{s}$ = 7\TeV is presented.
The results are based on data corresponding to an integrated luminosity of 2.3\fbinv
collected by the CMS detector at the LHC. Selected events are required to have one
isolated, high transverse momentum electron or muon, large missing
transverse energy, and hadronic jets, at least one of which must be
consistent with
having originated
from a b quark.
The measured cross section is
$158.1 \pm 2.1\stat \pm 10.2\syst \pm 3.5\lum\unit{pb}$, in agreement with standard model predictions.}

\hypersetup{%
pdfauthor={CMS Collaboration},%
pdftitle={Measurement of the t t-bar production cross section in pp collisions at sqrt(s) = 7 TeV with lepton + jets final states},%
pdfsubject={CMS},%
pdfkeywords={CMS, LHC, physics, top quark, cross section}}

\maketitle 

\section{Introduction}
Since the discovery of the top quark at the Fermilab Tevatron collider~\cite{top-discovery-cdf,top-discovery-d0},
considerable advances have been made in understanding its
production rates and decay properties in $\Pp\Pap$ collisions.
The advent of pp collisions at the Large Hadron Collider (LHC)~\cite{lhc} has
started a new phase of top quark physics,
and the first measurement at the higher center-of-mass energy of 7\TeV
was the top quark pair production cross section~\cite{top-xsec-cms,top-xsec-dil-cms,
top-xsec-atlas,top-xsec-dil-atlas}.
A precise measurement of the \ttbar cross section provides constraints
for QCD calculations presently available up to approximate
next-to-next-to-leading order (NNLO)~\cite{Aliev:2010zk,Langenfeld:2009wd, Kidonakis:2010dk,Ahrens}.
It is also important
for probing new physics processes that can manifest themselves as
an enhancement of the \ttbar production rate.

In this article, we present a precise measurement of
the \ttbar production cross section in pp collisions at $\sqrt{s}= 7\TeV$ utilizing a data set
corresponding to an integrated luminosity of 2.3\fbinv
recorded by the Compact Muon Solenoid (CMS) experiment at the LHC.

In the standard model (SM), top quarks are produced in pp collisions
predominantly via the strong interaction as \ttbar pairs, with each top quark
decaying almost exclusively into a W boson and a bottom quark.
In the analysis presented here, \ttbar events are identified in final states in which
one of the W bosons decays into a quark pair
and the other into a charged lepton
(electron or muon) and a neutrino, resulting in events that contain an electron or a muon,
a neutrino, and four hadronic jets, two of which result from hadronization of the b and
$\cPaqb$ quarks (b-jets). In order to improve the purity of the \ttbar candidate event sample,
we employ b-tagging algorithms, which are optimized for identification of b-jets. Decays of W bosons
into $\tau$ leptons are not specifically selected in this analysis, albeit some events enter the event sample
due to leptonic decays of the $\tau$.

The technique for measuring the \ttbar cross section from the candidate event sample consists of a
simultaneous profile likelihood fit to the distribution of invariant
masses of particles belonging to identified displaced vertices. These fits are performed as a function of the jet
and b-tag multiplicities in the event.
The method is similar to the one that was used in a previous CMS measurement~\cite{top-xsec-cms}, though a larger data sample is now studied.
Several alternative methods have been employed.
In one of these, we perform an inclusive measurement of \ttbar production cross section
without b-jet identification requirement, while others incorporate
different b-tagging algorithms.

\section{The CMS Detector}
The characteristic feature of the CMS detector is a superconducting solenoid
of 6\unit{m} in diameter, providing an axial magnetic field of 3.8\unit{T}.
Charged particle trajectories are measured by the silicon pixel and strip
subdetectors, covering $0 < \phi < 2\pi$ in azimuth and $|\eta| <~2.5$,
where the pseudorapidity $\eta$ is defined as
$\eta = - \ln[\tan\theta/2]$, with $\theta$ being the polar
angle of the trajectory of the particle with respect to the counterclockwise-beam
direction.
Within the field volume, the silicon detectors are surrounded by
a crystal electromagnetic calorimeter and a brass/scintillator hadron
calorimeter that provide high resolution energy measurement
of photons, electrons and hadronic jets.
Muon detection systems are located outside of the solenoid
and embedded in the steel return yoke. They provide muon
detection in the range $|\eta| < 2.4$.
 A two-tier trigger system selects the most interesting pp collision
events for use in physics analysis.
A detailed description of the CMS detector can be found in Ref.~\cite{cms}.

\section{Event selection
\label{sec:EventSelection}}
The sample of candidate \ttbar events is collected using
dedicated triggers, which require either a muon with transverse momentum
($\pt$) larger than 30\GeV or a high-$\pt$ electron. The criteria for the
electron trigger evolved during the course of data-taking in order to
maintain a reasonable trigger rate as the instantaneous luminosity of the
LHC increased.
For the initial data set, corresponding to an integrated luminosity 0.9\fbinv,
the threshold on the \pt of electron candidates varied between 27 and 32\GeV.
For the second part of the data set (1.4\fbinv)
the trigger required the presence of an electron with $\pt > 25$\GeV and at least
three hadronic jets with  $\pt > 30$\GeV.

The recorded events are reconstructed using the CMS particle-flow algorithm~\cite{ipartf}, which
categorizes observable particles into muons, electrons, photons, charged and neutral hadrons.
Energy calibration is performed separately for each particle type.
In the offline selection, muons are required to have a good-quality track
with $\pt > 35\GeV$ and $|\eta| <2.1$,  and the reconstructed tracks in the
silicon tracker are consistent with the track information from the muon systems~\cite{CMS_mu}.
Electrons are identified using a combination of the shower shape information
in electromagnetic calorimeter and track-cluster matching~\cite{CMS_e}, and are
required to have $\pt > 35\GeV$ and $|\eta| < 2.5$.
Electron candidates in the
transition region between the barrel and forward electromagnetic calorimeters,
$1.44 < |\eta| < 1.57$,
are not used for the measurement.
We also reject electrons coming
from photon conversions~\cite{CMS_e}.

Since the lepton from a W decay is expected to be isolated from other activity
in the event, we apply isolation requirements. The relative isolation is defined as
$I_{\text{rel}} =\big(\sum \ET^{\text{charged}} + \sum \ET^{\text{photon}} + \sum \ET^{\text{neutral}}\big)/\pt$,
where $\pt$ is the lepton transverse momentum, and
$\ET^{\text{charged}}$, $\ET^{\text{photon}}$, and $\ET^{\text{neutral}}$ are transverse energies
of the charged particles, the reconstructed photons, and the neutral particles not identified as photons.
The sum of the transverse energies is computed
in a cone of size
$\Delta R =\sqrt{(\Delta\phi)^2+(\Delta\eta)^2} = 0.3$ around the lepton direction, excluding the
lepton candidate itself.
We require $I_{\text{rel}}$ to be less
than 0.125 for muons and 0.10 for electrons.

The signal events are required to have only one electron or muon whose
origin is consistent with the reconstructed primary pp interaction vertex~\cite{CMS_vertex},
defined as the vertex with the largest value for the scalar
sum of the $\pt$ of the associated tracks.
Events with an additional electron or muon
candidate that satisfies less strict lepton identification requirements are vetoed.

Jets are reconstructed using the particle-flow algorithm and are clustered using the 
anti-\kt jet technique~\cite{ktalg} with a distance parameter of 0.5, as implemented in
\textsc{FastJet} v2.4.2~\cite{fastjet1, fastjet2}. In order to account for extra
activity within a jet cone from multiple pp interactions per beam crossing,
referred to as a pileup, jet energies are corrected for charged
hadrons that originate from a vertex other
than the primary one, and for the amount of pileup expected in the jet area
from neutral jet constituents. Jet energies are also corrected for non-linearities
due to different responses in the endcap and barrel calorimeters, and differences
between true and simulated calorimeter responses~\cite{jec_11}.
Each jet is required to have a
transverse momentum $\pt > 35\GeV$ and $|\eta| < 2.4$.
We select events with at least one jet, or at least three jets for
events collected with the electron + jets trigger.
To reduce background processes,
we require at least one of the jets to be identified as a b-jet by a displaced
secondary vertex algorithm known as \textit{Simple Secondary Vertex High Efficiency}~\cite{CMS_btag}
with a medium working point.
The algorithm has a b-tag efficiency of 55\% and a light parton (u, d, s, g)
mistag rate of 1.5\%.

In addition, events are required to have a significant amount of missing transverse
energy $(\ETslash)$ as evidence of a neutrino from the W boson decay.
This is defined as the magnitude of the negative vector sum of the transverse momenta
of all of the objects found by the particle-flow algorithm.
We require $\ETslash > 20\GeV$ for both the electron + jets and muon + jets channels.

\section{Signal and Background Modeling\label{sec:sigback}}
Pair production of top quarks is modeled using the \textsc{MadGraph v5.1.1}~\cite{Alwall:2011}
Monte Carlo (MC) event generator, assuming the mass of the top quark m$_\cPqt = 172.5$\GeV.
The top quark pairs are generated with up to three additional hard jets
using \PYTHIA~v6.424 with tune Z2~\cite{Sjostrand:2006za}
to model parton-showering (PS),
and the shower matching is performed using the
\kt-MLM prescription~\cite{Alwall:2011}.
The generated events are further passed through the full CMS detector simulation based on
\GEANTfour~\cite{Geant4}. The presence of pileup is incorporated by simulating additional
interactions with a multiplicity matching that observed in data.

Leptonically decaying W + jets events constitute by far the largest background.
These together with Z + jets events are also generated using \MADGRAPH
with up to four jets subject to the matrix-element (ME) description.
The W + jets events are generated inclusively with respect to jet flavor.
Reconstructed jets are further matched to partons in the simulation, and
the W + bottom quark and W + charm quark components are separated
from the W + light-flavor (u, d, s, and gluon) component
based on the parton flavor.

Other backgrounds include single-top-quark production,
simulated with \textsc{powheg v1.0}~\cite{powheg1, powheg2, powheg3}, QCD multijet simulated with \PYTHIA,
and photon + jet events, which constitute a background for the electron + jets
channel, generated by \MADGRAPH. The set of parton distribution functions
used by \MADGRAPH is
CTEQ6L1~\cite{cteq_2010},
while  \POWHEG and \PYTHIA
use CTEQ6M~\cite{cteq_2010}.

The W and Drell--Yan production processes are normalized based on NNLO cross sections,
determined using \textsc{fewz}~\cite{fewz}. They correspond to
$\sigma_{\PW\to\ell \nu} = 31.3 \pm 1.6\unit{nb}$ and
$\sigma_{\cPZ/\gamma^*\to\ell \ell} = 3048\pm 132\unit{pb}$, where
for the Drell--Yan production the invariant mass of two leptons ($\ell = \Pe$ or $\mu$) is
greater than 50\GeV.
The single-top-quark $t$-channel production is normalized to the recent CMS measurement of $\sigma_{\cPqt} = 67.2 \pm 6.1\unit{pb}$~\cite{cms_SingleTop}.
The single-top-quark associated production $(\cPqt\PW)$ is normalized to the approximate NNLO cross section
$\sigma_{\cPqt\PW} = 15.7 \pm 1.2\unit{pb}$~\cite{Kidonakis:2010tW}, and the $s$-channel is normalized to the
next-to-next-to-leading-logarithm prediction of $\sigma_{\cPqt} = 4.6 \pm 0.2\unit{pb}$~\cite{NNLLtop}.

The QCD multijet normalization is obtained by fitting SM contributions to the full
$\ETslash$ distribution in data, though only the yield of QCD multijet events with $\ETslash > 20$\GeV enters the normalization. For the electron + jets
channel, the QCD multijet background distributions are obtained from MC, and for the muon + jets
channel, they are obtained from a background-enriched data sample defined as
$I_{\text{rel}} > 0.125$ and $\ETslash <20\GeV$.

\section{Cross Section Measurement}
The \ttbar cross section measurement is performed using a maximum profile likelihood fit to the
number of reconstructed jets ($N_{\text{jet}}$), the number of b-tagged jets ($N_{\text{tag}}$), and the secondary vertex
mass (SVM) distribution in the data. We consider ten event subsamples with $N_{\text{jet}}$
values of 1--4 and ${\ge}5 $, and $N_{\text{tag}}$ values of  1 and ${\ge}2$.
The SVM is defined as the mass of the sum of four-vectors of the tracks associated to the
secondary vertex with an assumption that all particles have the pion mass.
For events with two b-tagged jets,
SVM corresponds to the highest-$\pt$ b-tagged jet.
The SVM distribution yields a good discrimination between the contributions from light- and heavy-flavor quark
production~\cite{top-xsec-cms}. The results are obtained by
maximizing a binned Poisson likelihood that incorporates contributions from \ttbar,
W + jets, Z + jets, single-top-quark, and QCD multijet production processes.
Performing a simultaneous fit across different jet and b-tag multiplicity bins, including
regions dominated by background events, constrains the background contributions,
resulting in a more precise measurement of the \ttbar production cross section.

The W + jets, Z + jets, and single-top-quark background processes are initially
normalized to the expected event yields according to their theoretical cross
sections. The QCD and photon + jets normalizations
are evaluated as described above individually in each $N_{\text{jet}}$ and $N_{\text{tag}}$ subsample, for both channels.
These background normalizations are
the initial values that enter the profile likelihood fit. The cross section measurement is
performed by fitting to the data to obtain corrections to these initial values.
The W + jets backgrounds are split into W + b jets, W + c jets,
and W + light-flavor (LF) sub-samples, with all three components free
in the fit. During the likelihood maximization, the normalizations of each of these
components are extracted. The normalizations of the \ttbar and W + jets contributions
are allowed to float freely.
The contributions from small backgrounds, QCD multijet
and Z + jets, are conservatively
constrained with Gaussian uncertainties of 100\% and 30\% of their
expected event yields, respectively.
The single-top-quark contribution is constrained with
an uncertainty of 10\%~\cite{cms_SingleTop}.

The expected event yield for each background component, per $N_{\text{jet}}$ and $N_{\text{tag}}$,
is also a function of other parameters, such as the jet energy scale (JES),
the b-tagging efficiency and the mistag rate. In addition, the $N_{\text{jet}}$ spectrum is affected
by the choice of the renormalization and factorization ($Q^2$) scales.
For the W + jets simulation we use a dynamical mass scale of
$(m_{\PW})^2 + (\sum \pt^{\text{jet}})^2$, where $m_{\PW}$ is the mass of the W and $\sum \pt^{\text{jet}}$ is the
sum of the transverse momenta from the jets in the event.
The magnitude of the scale is allowed to vary in the fit
by incorporating an effective parameter $c_{Q^2}$ into the likelihood
with initial value 1.0, and which is allowed to vary between 0.5 and 2.0.
The profile likelihood maximization
provides simultaneous measurements of each of these parameters, background contributions
and the \ttbar cross section.

There are alternative control samples to estimate the JES, the b-tagging efficiency and the
mistag rate. The JES
uncertainty is measured in control samples to be approximately 3\%~\cite{jec_11}, and this determines
a Gaussian constraint on this parameter in the likelihood.
To account for differences between simulation and data in the b-tagging efficiencies and the mistag rates, we weight the tagged jets in the
simulation up or down by a data-to-simulation scale factor.
The b-tagging efficiency and the mistag rate scale factors are constrained to be
$1.0 \pm 0.1$ in the fit, where 10\% is the uncertainty in the b-tagging efficiency and
the mistag rate~\cite{CMS_btag}.

The systematic uncertainties related to JES, $Q^2$ scale,
b-tagging and mistag scale factors are included as nuisance parameters in the profile
likelihood fit. Other systematic uncertainties are not directly
included in the profile likelihood and taken as additional systematic uncertainties
outside of the fit result and are described below.

The efficiencies for triggering, reconstructing, and identifying isolated leptons are
determined using $\cPZ\to\Pe\Pe$ and $\cPZ\to\mu\mu$ samples of events, and found to be
very similar in the data and simulation. We correct for small differences observed,
and account for an additional systematic uncertainty of 3\% on these values.
The unclustered energy in the detector results in an additional resolution
uncertainty of less than 1\% on the $\ETslash$ scale. The difference in jet energy resolution
determined
in simulation and data results in an uncertainty of less than 1\%.

The theoretical uncertainties in modeling of \ttbar production are evaluated from
dedicated simulated event samples by varying the theoretical parameters
of interest around their nominal values. Such variations are used to construct alternative distributions,
from which simulated events can be generated.
For each variation, 4000 pseudo-experiments are generated and fitted with the standard configuration.
The mean bias of the fitted \ttbar cross section is taken as the size of the systematic uncertainty
due to the source under study.
These include differences in the \ttbar signal due to renormalization
and factorization scales (4\%),  the scale for the ME partons
to PS matching scheme (2\%),
pileup modeling in simulation (less than 1\%), and the parton distribution
function model (less than 1\%). The total uncertainty for the \ttbar modeling, when adding the above
uncertainties in quadrature, is 5.0\%.

The systematic uncertainty on the SVM shape is also considered.
We have studied several effects, which include pixel resolution and jet-track-association modeling, as well as pileup dependence. These have a negligible effect on the SVM shapes of
\ttbar, single-top-quark, W and Z + jets events. The uncertainty on the SVM shape from QCD multijet background is obtained as follows.
For the electron + jets channel
we generate
pseudo-experiments based on the default and alternative QCD shapes obtained from simulation. To increase the statistical accuracy of the QCD multijet background, the default shape
employed in the fit is taken from events with relaxed requirements on the electron isolation
and identification, and no $\ETslash$ requirement imposed.
The alternative shape is obtained from the region corresponding to the event selection
used in the \ttbar cross section measurement.
For the muon + jets channel,
the statistical fluctuations in the normalization for the $\ETslash$ distributions obtained from muon
non-isolated ($I_{\text{rel}} > 0.125$) and isolated ($I_{\text{rel}} < 0.125$) regions are taken as the
systematic uncertainty. The integrated luminosity of the event sample is determined with an
uncertainty of 2.2\%~\cite{Lumi}.

The list of systematic
uncertainties is summarized in Table~\ref{tab:sys_summary}.
These include both the uncertainties related
to the nuisance parameters in the likelihood fit and
the additional uncertainties evaluated from alternative distributions as described above.
The individual systematic uncertainties related to
the nuisance parameters in the fit
are shown for illustrative purposes only. These are obtained as follows. First, the total fit uncertainty
is evaluated when the parameter of interest is fixed in the fit. Then this uncertainty is subtracted
in quadrature from the total fit uncertainty when all parameters are varied in the fit.
Since the treatment of the $Q^2$ uncertainty in the likelihood fit is dependent
on parametrization, we also performed the cross-check with the $Q^2$ uncertainty
treated outside of the fit, and obtained consistent results.
The combined systematic uncertainty of the measurement is 6.5\%,
taking into account the correlations between the nuisance parameters.

\begin{table*}[htb]
\topcaption[Summary Systematics Table]{\label{tab:sys_summary}
List of systematic uncertainties for the electron + jets, muon + jets, and the combined analysis.
Due to the correlation between the fit parameters, the combined number is not the root of the quadratic sum
of the contributions.}
\begin{center}
\begin{tabular}{cccc}
\hline
Source       & Electron & Muon    & Combined \\
             & channel  & channel & analysis \\
\hline
 \multicolumn{1}{c}{Quantity}  &  \multicolumn{3}{c}{Uncertainty\,(\%)}  \\
\hline
Lepton ID/reconstruction/trigger efficiency    & 3.0  & 2.0  & 3.0   \\
$\ETslash$ resolution due to unclustered energy      & 0.9  & 0.3  & 0.8   \\
Jet energy resolution                          & 0.5  & 0.5  & 0.6   \\
\ttbar + jets renorm./fact. scales             & 3.5  & 4.3  & 4.3   \\
\ttbar + jets ME to PS matching                & 2.2  & 1.8  & 2.2   \\
Pileup                                         & 0.5  & 0.2  & 0.6   \\
Parton distribution function choice            & 0.3  & 0.3  & 0.3   \\
QCD multijet SVM distribution                 & 1.4  & 0.4  & 0.6   \\ \hline
Subtotal                     & 5.5  & 5.1  & 5.9   \\ \hline
\multicolumn{1}{c}{Nuisance parameter}  & \multicolumn{3}{c}{Uncertainty\,(\%)}  \\
\hline
Jet energy scale                              & 3.9  & 3.0  & 2.4  \\
b-tagging efficiency and mistag rate      & 3.5  & 2.8  & 2.1  \\
W + jets renorm./fact. scale                  & 1.6  & 1.5  & 1.6  \\
\hline
Total systematic uncertainty                  & 7.0  & 6.2  & 6.5 \\
\hline
\end{tabular}
\end{center}
\end{table*}

The measurement is performed separately for the electron + jets and muon + jets channels,
as well as simultaneously for both channels, yielding
\begin{description}
\item{\textbf{Electron + jets}}
\begin{equation}
\sigma_{\ttbar} = 160.6  \pm 3.2\stat \pm 11.2\syst \pm 3.5\lum\unit{pb},
\end{equation}
\item{\textbf{Muon + jets}}
\begin{equation}
\sigma_{\ttbar} = 164.2   \pm 2.8\stat \pm 10.1\syst \pm 3.6\lum\unit{pb},\ \text{and}
\end{equation}
\item{\textbf{Combined}}
\begin{equation}
\sigma_{\ttbar} = 158.1   \pm 2.1\stat \pm 10.2\syst \pm 3.5\lum\unit{pb}.
\end{equation}
\end{description}
\begin{figure}[h!tb]
\centering
\includegraphics[width=\cmsFigWidth]{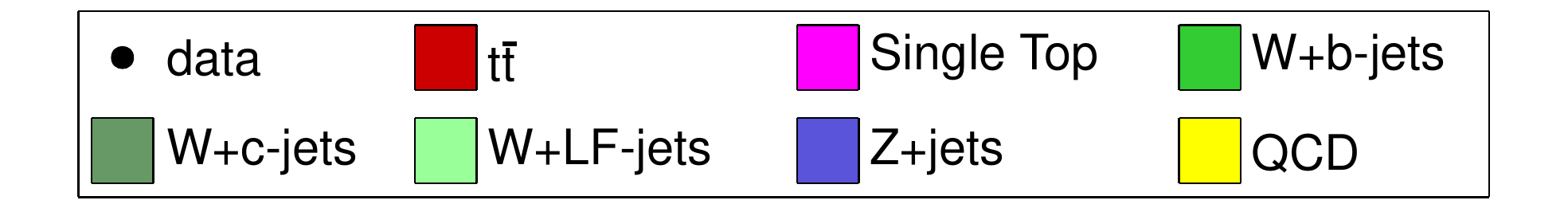}
\includegraphics[width=\cmsFigWidth]{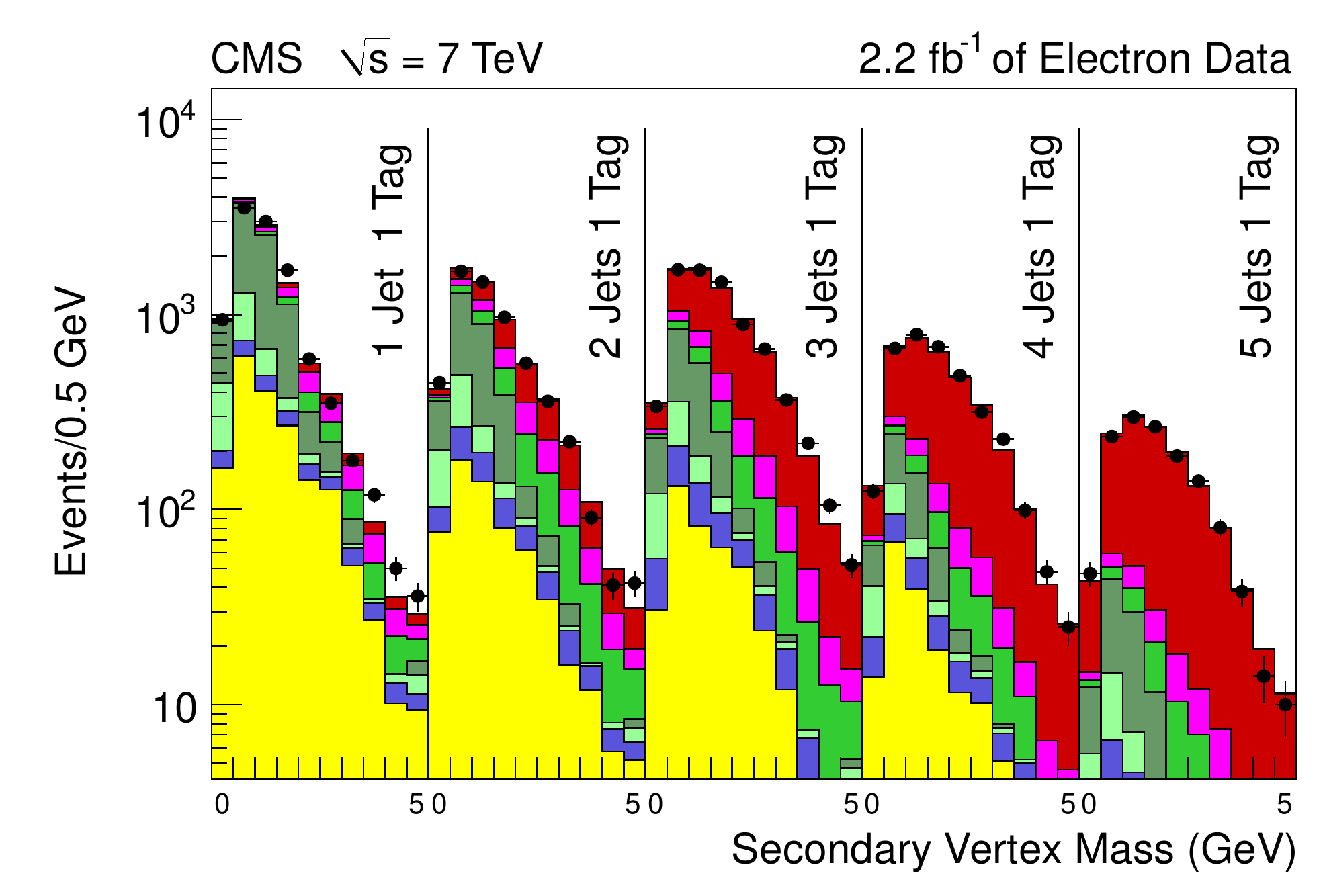}
\includegraphics[width=\cmsFigWidth]{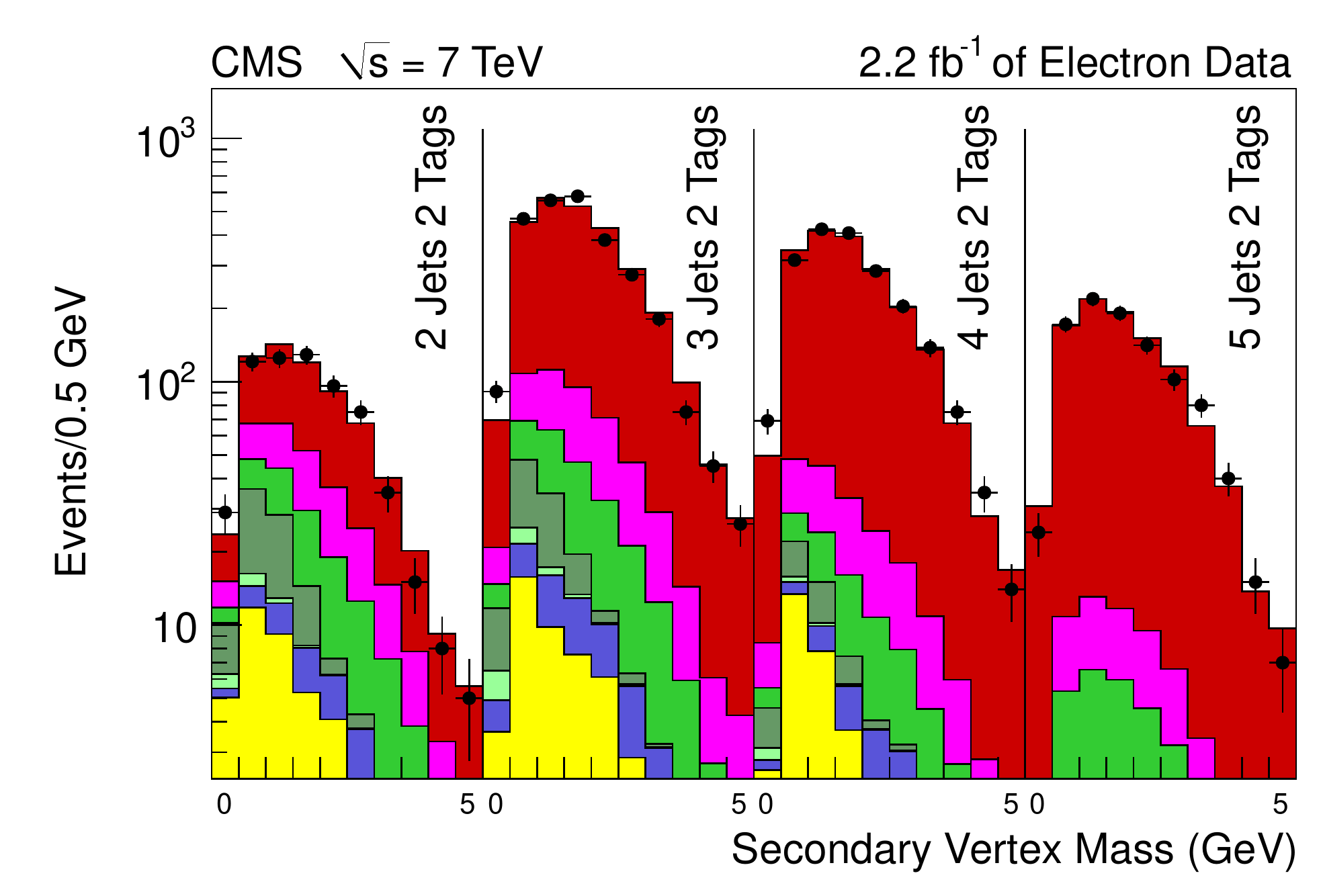}
 \caption[Results of the combined fit.]{Results of the combined fit for the electron + jets channel,
 for single b-tag events (top panel), and for $\geq$2b-tag events (bottom panel).
The distributions within
each panel correspond to events with 1-, 2-, 3-, 4-, and $\ge$5-jets, respectively. The error
bars indicate statistical uncertainties only.}
\label{fig:ele_tagged_2fb}
\end{figure}

\begin{figure}[h!tb]
\centering
\includegraphics[width=\cmsFigWidth]{legend_f18_ncol4_lessWhite}
\includegraphics[width=\cmsFigWidth]{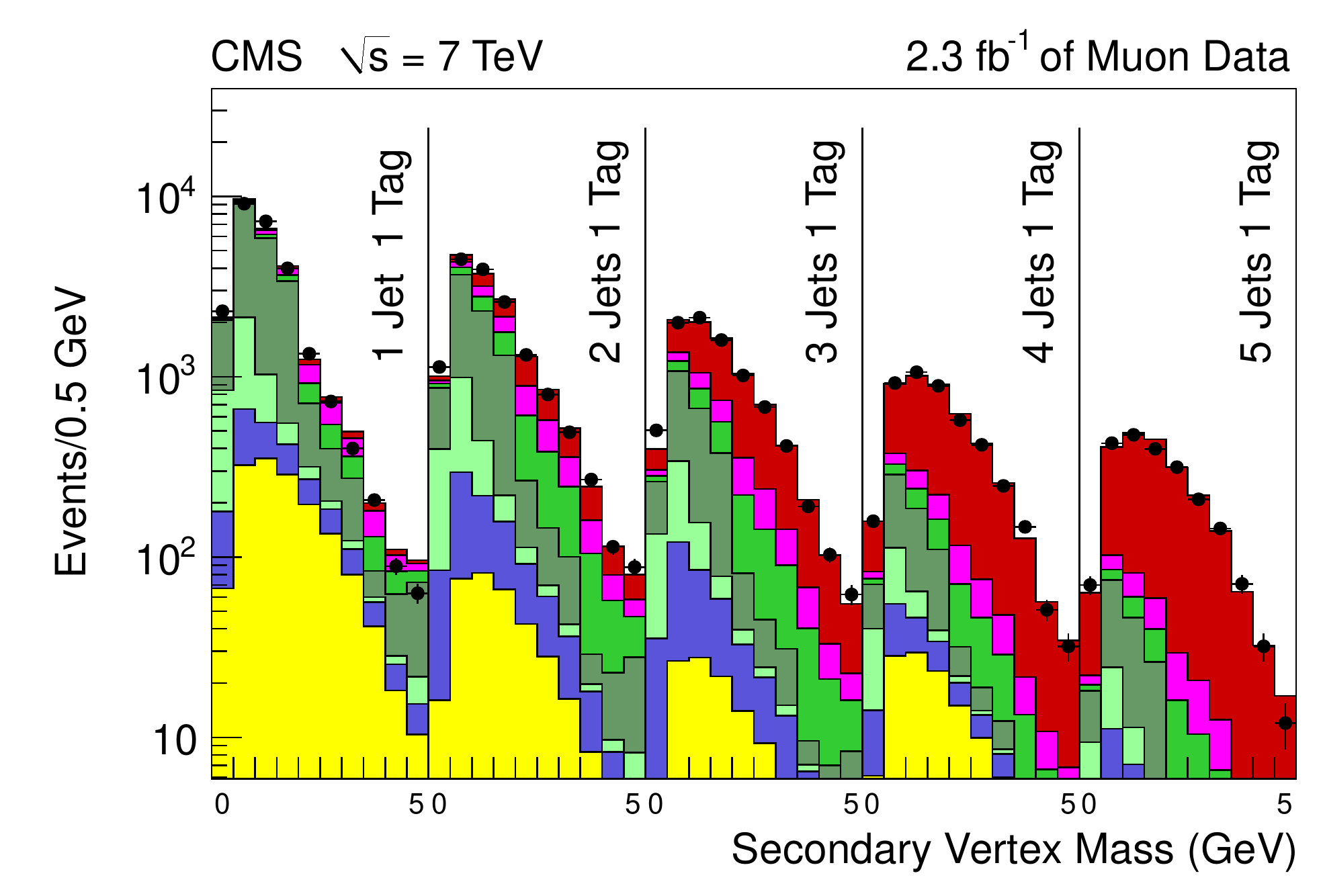}
\includegraphics[width=\cmsFigWidth]{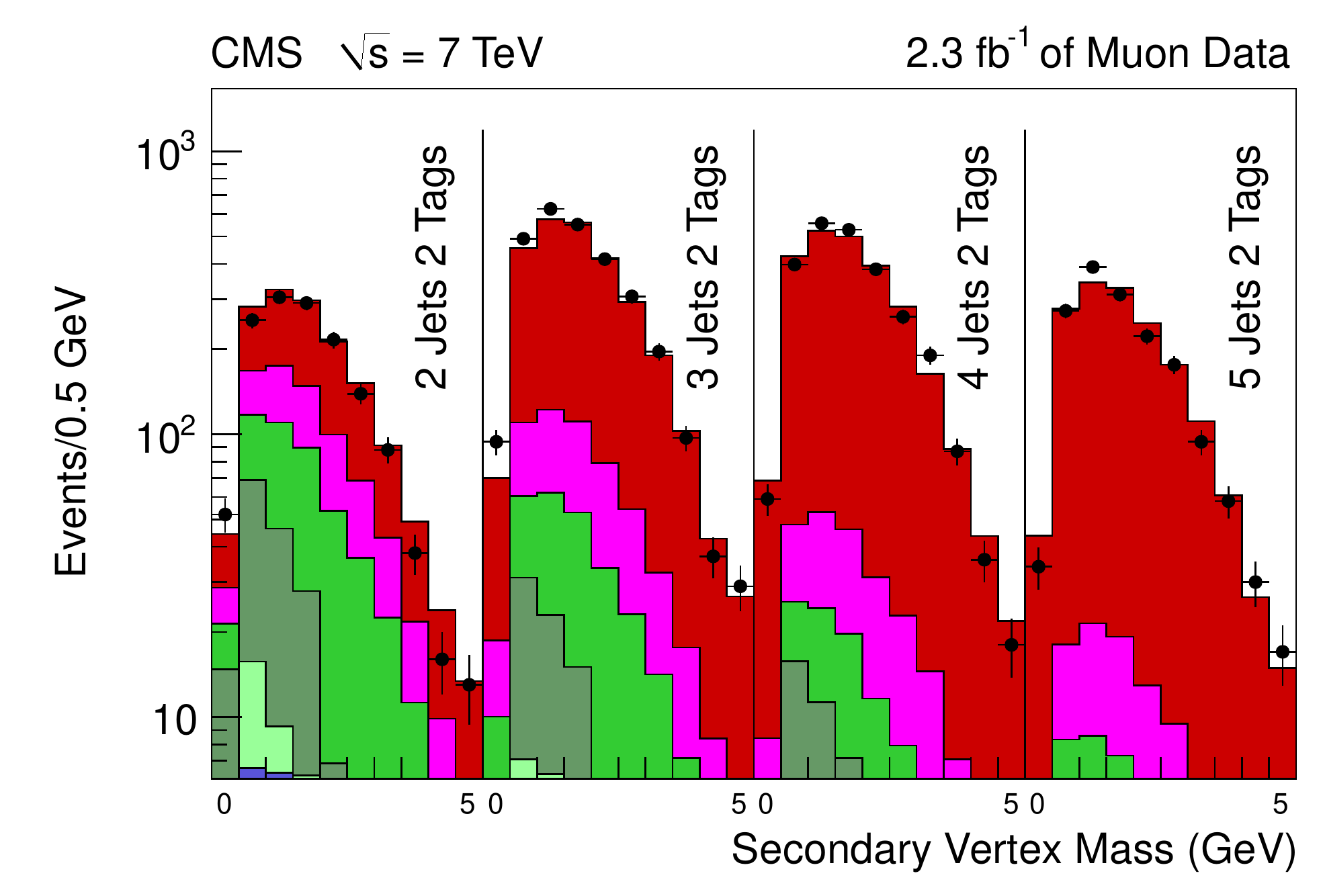}
\caption[Results of the combined fit.]{
Results of the combined fit for the muon + jets channel,
 for single b-tag events (top panel), and for $\geq$2b-tag events (bottom panel).
The distributions within
each panel correspond to events with 1-, \mbox{2-}, 3-, 4-, and $\ge$5-jets, respectively. The error
bars indicate statistical uncertainties only.
\label{fig:muo_tagged_2fb}}
\end{figure}

The comparison of the corresponding observed and fitted SVM distributions
is shown in Figures~\ref{fig:ele_tagged_2fb} and~\ref{fig:muo_tagged_2fb}.
As a by-product, the fit
provides the size
of contributions from
the SM processes
that are backgrounds to \ttbar production, as well as in-situ evaluations
of other parameters varied
in the profile likelihood fit, such as the b-tagging efficiency and the JES correction
factor (on top of the standard jet corrections).
The results of the combined fit, as well as the results of the fits performed in the electron + jets and
muon + jets samples separately, are listed in Table~\ref{table:side_by_side}, with correlations among parameters shown in Table~\ref{table:corr_combined_shyfts_12}. The \ttbar cross section is
given in pb, while the contributions from other standard model processes are quoted as scale factors with respect to their theoretical predictions described above. These measured scale factors do not account for a full treatment of the systematic uncertainties and hence are strictly valid only in the context of the fit presented in this paper.
The b-tagging scale factor defined as the ratio of the b-tagging efficiencies in data and simulation
is determined to be $96 \pm 1$\%, consistent between the electron + jets and muon + jets channels.
The JES correction factor is found to be
$100.4 \pm 1.6$\%
and
$98.1 \pm 1.2$\%
in the electron + jets and muon + jets channels, respectively, yielding
$100.2 \pm 1.0$\%
in the combined fit.

The W + c jets contribution in the data is found to be
larger than SM predictions, both in the electron + jets and
muon + jets channels. This contribution includes single charm and double charm production,  which are both present in the selected events.
The W + b jets contribution in the data is also found to be slightly higher than in the simulation. The W + LF jets scale factor in the electron channel is significantly lower than in the muon case. This is because of the presence of a much larger QCD multijet contribution in the electron sample, and its large correlation with the W + LF jets component. The combined W + LF jets/QCD multijet scale factors for muons and electrons are in agreement, being $0.84 \pm 0.09\%$ and $0.71 \pm 0.07\%$, respectively.

Because of large correlated uncertainties
and due to differences in the correlation matrix for
the electron + jets and muon + jets channels, which
result from different QCD multi jet contributions in the two channels,
the result of the combined fit resides
outside of the individual 
electron + jets and muon + jets
measurements.
The correlation matrix for the combined fit is given in
Table~\ref{table:corr_combined_shyfts_12}.
Using 4000 alternative data sets constructed from the simulated events
we determine that the combined cross section lies
between the individual channel results only in 60\% of the cases.
For the combined fit we have seven out of ten parameters that are common to both channels,
residing outside of the ${\pm}1 \sigma$ interval between individual electron + jets and muon + jets
measurements. Using simulated events we determine this to occur in 10\% of the cases.

\begin{table*}[hbt]
\topcaption{\label{table:side_by_side}
{Results of the fit to the combined electron + jets and  muon + jets sample, and each
channel individually.
The contributions from the background processes are quoted as scale factors
with respect to their theoretical predictions.
The scale factors do not account for a full treatment of the systematic uncertainties and are strictly valid only in the context of the fit.
For brevity, the QCD parameters are not shown.
}}
\begin{center}
\begin{tabular}{ l  r  r  r }
\hline
         Fit parameters  & Electron + jets                 & Muon + jets                   & Combined \\
\hline
                          $\sigma_{\ttbar}$ (pb)	 & $ 160.6 \pm 6.6 $   &  $  164.2 \pm 5.5  $  &  $  158.1 \pm 4.1 $ \\

\hline
              Single top & $  1.05  \pm 0.10 $   &  $   1.08 \pm 0.10 $  &  $   1.17 \pm 0.10 $ \\
             W + b jets  & $  1.19 \pm 0.35 $   &  $   0.95 \pm 0.18 $  &  $   1.28 \pm 0.16 $ \\
             W + c jets  & $  1.54 \pm 0.15 $   &  $   1.48 \pm 0.05 $  &  $   1.55 \pm 0.04 $ \\
             W + LF jets & $  0.20 \pm 0.08 $   &  $   0.57 \pm 0.07 $  &  $   0.52 \pm 0.06 $ \\
           Z + jets 	 & $  1.13 \pm 0.29 $   &  $   1.08 \pm 0.29 $  &  $   1.43 \pm 0.29 $ \\
\hline
                $c_{Q^2}$   &  $   1.02 \pm 0.16 $  &  $   0.94 \pm 0.06 $ &  $   1.05 \pm 0.05  $ \\

              b-tag      &  $   0.95 \pm 0.01$  &  $  0.97 \pm 0.01$ &  $  0.96 \pm 0.01 $ \\
              JES  	 &  $   1.00 \pm 0.02$  &  $  0.98 \pm 0.01 $ &  $  1.00 \pm 0.01 $ \\
              Mistag 	 &  $   1.00 \pm 0.10 $  &  $  1.00 \pm 0.10 $ &  $  1.00 \pm 0.10 $ \\
\hline
\end{tabular}
\end{center}
\end{table*}

\begin{table*}[hbt]
\topcaption{\label{table:corr_combined_shyfts_12}
{Correlation matrix of the combined fit to the electron + jets plus
  muon + jets samples. Only non-QCD parameters are shown.}}
\begin{center}
\begin{tabular}{ l  rrrrrrrrrr}
\hline
 & \ttbar & Single t & W + b & W + c & W + LF & Z + jets & $c_{Q^2}$ & b-tag & JES & Mistag \\ \hline
       \ttbar &$    1.00 $&$  -0.13 $&$  -0.48 $&$   0.33 $&$   0.03 $&$   0.07 $&$  -0.07 $&$  -0.70 $&$  -0.81 $&$   0.00 $ \\
     Single t &$   -0.13 $&$   1.00 $&$  -0.52 $&$   0.04 $&$   0.03 $&$  -0.03 $&$   0.06 $&$  -0.08 $&$   0.09 $&$  -0.00 $ \\
        W + b &$   -0.48 $&$  -0.52 $&$   1.00 $&$   0.05 $&$   0.13 $&$  -0.16 $&$   0.27 $&$   0.26 $&$   0.42 $&$  -0.02 $ \\
       W + c  &$    0.33 $&$   0.04 $&$   0.05 $&$   1.00 $&$   0.01 $&$   0.15 $&$   0.71 $&$  -0.38 $&$  -0.26 $&$  -0.02 $ \\
       W + LF &$    0.03 $&$   0.03 $&$   0.13 $&$   0.01 $&$   1.00 $&$  -0.19 $&$   0.21 $&$  -0.03 $&$  -0.05 $&$  -0.83 $ \\
     Z + jets &$    0.07 $&$  -0.03 $&$  -0.16 $&$   0.15 $&$  -0.19 $&$   1.00 $&$   0.23 $&$  -0.01 $&$  -0.10 $&$   0.01 $ \\
    $c_{Q^2}$ &$   -0.07 $&$   0.06 $&$   0.27 $&$   0.71 $&$   0.21 $&$   0.23 $&$   1.00 $&$  -0.02 $&$   0.15 $&$  -0.02 $ \\
        b-tag &$   -0.70 $&$  -0.08 $&$   0.26 $&$  -0.38 $&$  -0.03 $&$  -0.01 $&$  -0.02 $&$   1.00 $&$   0.43 $&$  -0.02 $ \\
          JES &$   -0.81 $&$   0.09 $&$   0.42 $&$  -0.26 $&$  -0.05 $&$  -0.10 $&$   0.15 $&$   0.43 $&$   1.00 $&$   0.01 $ \\
       Mistag &$    0.00 $&$  -0.00 $&$  -0.02 $&$  -0.02 $&$  -0.83 $&$   0.01 $&$  -0.02 $&$  -0.02 $&$   0.01 $&$   1.00 $ \\
\hline
\end{tabular}
\end{center}
\end{table*}

The \ttbar cross section is measured assuming a value of the top quark mass $m_{\cPqt} = 172.5$\GeV.
The measured cross section of \ttbar production has a dependence on $m_\cPqt$, which is
evaluated using dedicated MC samples and can be parameterized in the range of 160--185\GeV as
\begin{equation}
\sigma_{\ttbar} = 158.1\unit{pb} - (m_{\cPqt} -172.5\GeV ) \times (1.14 \pm 0.18\unit{pb}/\GeVns).
\end{equation}

\section{Alternative Analyses}\label{sec:crosscheck}

In addition to the main result,
we have performed several alternative analyses
in the electron + jets and the muon + jets channels
using different
event selections and
different methods to
suppress background contributions and measure the \ttbar cross section.
One analysis does not rely on b-tagging, a second one makes
use of the kinematical information from the top quark decays, and a
third one relies on a data-based estimate of the dominant background.

The analysis without relying on use of the b-tagging algorithms
considers the
data set
corresponding to
4.6\fbinv (4.9\fbinv) in the electron (muon) + jets
channel.
The selected events are required to
have an electron with $\pt > 35$\GeV or a muon with $\pt > 26$\GeV, and
at least 4 jets
with $\pt > 30$\GeV. No missing transverse energy requirement is imposed.

The cross section is measured using a binned log-likelihood fit to the mass of the three-jet combination with the highest $\pt$ in the event ($M_3$).
The \ttbar, W/Z + jets, and  QCD multijet components are unconstrained during
the fit, with the QCD multijet contributions in the electron + jets and muon + jets
channels treated independently. The single-top-quark normalization is constrained to within
30\% of its theoretical value.

The \ttbar, single-top-quark, and W/Z + jets processes are modeled using the simulation,
while
the QCD multijet contribution is estimated from data using
a side-band region
with the relative isolation of the lepton greater than 0.25.
Signal events, as well as W + jets events, are heavily suppressed by this selection,
and subtracted based on
simulation. The shape of the subtracted QCD multijet contribution
is used for the fit in the signal region, since the $M_3$ distribution of QCD events
does not depend on the relative isolation of the lepton in the event.

\begin{figure}[htbp]
 \centering
 \includegraphics[width=0.5\textwidth]{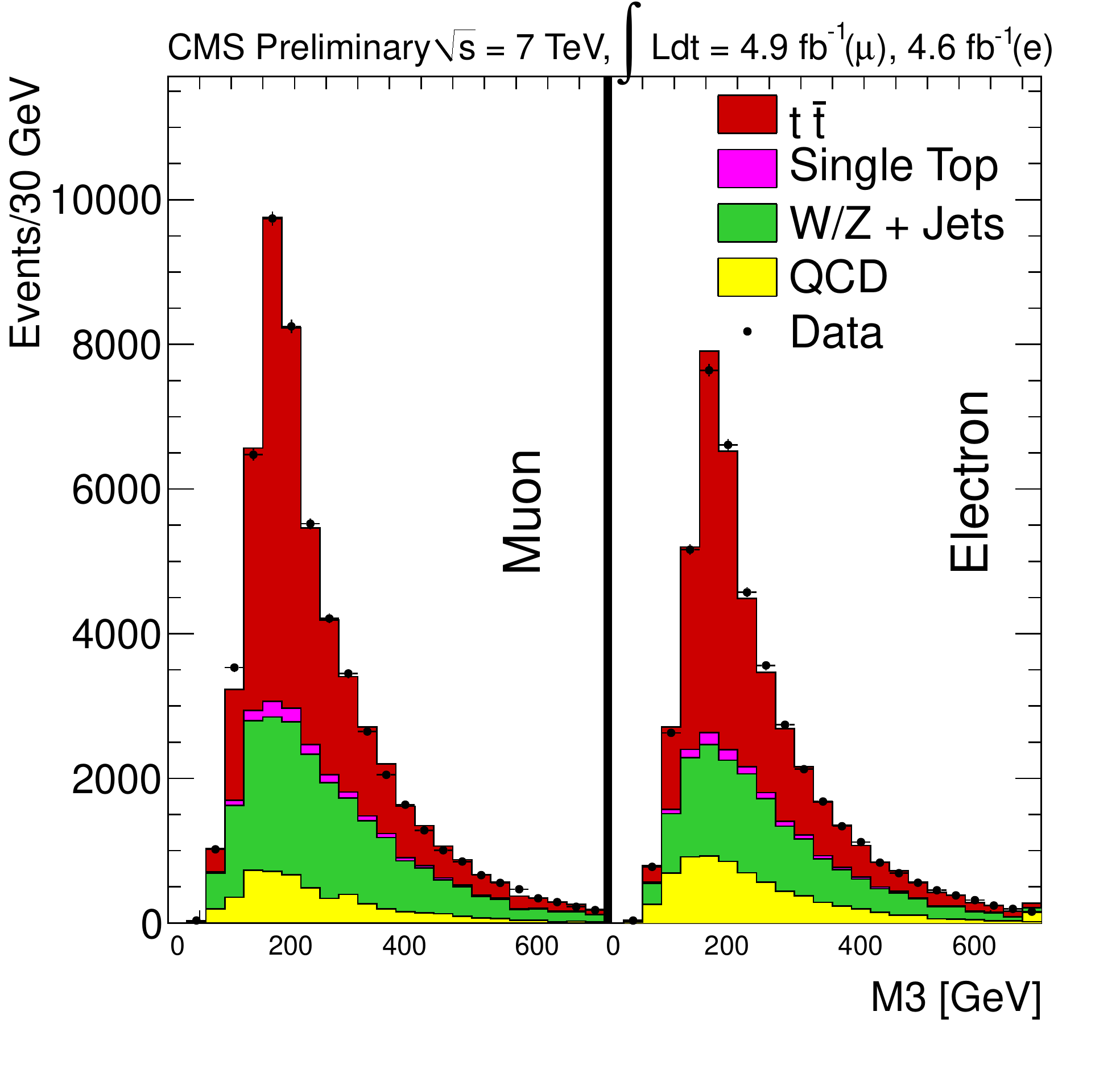}
 \caption{
 The mass of the three-jet combination
 with the highest transverse momenta.}
\label{fig:emu_data_mlfit_res_AN11_443}
\end{figure}

The observed and the fitted $M_3$ distributions are shown in
Figure~\ref{fig:emu_data_mlfit_res_AN11_443}.
The dominant sources of systematic uncertainty are JES, ME to PS matching,
and the $\mathrm{Q^2}$ scale uncertainties. In the electron + jets channel the cross
section measurement yields
\begin{equation}
\sigma_{\ttbar} = 157.1 \pm 3.7\stat^{+17.2}_{-11.4}\syst\pm 3.5 \lum\unit{pb},
\end{equation}
in the muon + jets channel the cross section is measured as
\begin{equation}
\sigma_{\ttbar} = 161.6 \pm 3.5\stat^{+14.8}_{-22.1} \mathrm{~(syst.)}  \pm 3.6 \lum\unit{pb},
\end{equation}
The combined measurement in the electron + jets and muon + jets channels yields a cross section of
\begin{equation}
\sigma_{\ttbar} = 159.7 \pm 2.6 \stat^{+13.1}_{-14.7}\syst\pm 3.5 \lum\unit{pb}.
\end{equation}

Another measurement uses kinematic information from the leptonic top quark decay
$\cPqt\to\cPqb\PW \to \cPqb\ell \nu_\ell$, namely the mass of the two-particle system consisting
of a lepton and a jet associated with a b quark. The jet-to-parton assignment among the four
leading jets is
performed minimizing a least-squares residual based on the masses of the reconstructed W boson and hadronically decaying top quark
in $\cPqt\to\cPqb\PW \to \cPqb\cPq\cPaq'$.
The baseline event selection is similar to the
reference analysis, complemented with the requirement that the jet assigned to the leptonic
top quark decay is b-tagged using an algorithm based on measuring the significance of a
track impact parameter~\cite{CMS_btag}.
The technique is applied to the muon + jets data sample, corresponding to an integrated
luminosity 4.9\fbinv.
The result of this measurement is
$\sigma_{\ttbar} = 162.4 \pm 5.4\stat^{+7.5}_{-11.0}\syst\pm 3.6\lum\unit{pb}$,
where the dominant
systematic uncertainty is due to the JES.

Finally, the
third method does not rely on MC simulation for the W + jets background,
but exploits the W charge asymmetry~\cite{cms_WchargeAsym} in W + jets production at the LHC.
The shape of the lepton pseudo-rapidity distribution for the W + jets component is obtained
from the data by subtracting the observed distribution for $\ell^-$ from
the one corresponding to $\ell^+$.
The \ttbar cross section is measured by fitting a combination of signal
and background components to the observed lepton $|\eta|$ spectrum
using a data sample corresponding to an integrated luminosity of 0.9 (1.0)\fbinv
for electron (muon) + jets.
The \ttbar cross section is measured with a large expected uncertainty of 42\%
(23\%) in the electron (muon) + jets channel, and agrees with the
results of the other analyses.

\section{Summary}

The \ttbar production cross section measurement has been performed at
$\sqrt{s}=7\TeV$ using the data collected with the CMS detector and corresponding
to an integrated luminosity of 2.3\fbinv.

The \ttbar cross section is measured using a profile likelihood fit
to the number of reconstructed jets, the number of b-tagged jets, and the
secondary vertex mass distribution. The measured cross section for an assumed
top quark mass of 172.5\GeV is
\begin{linenomath}
\begin{equation}
\sigma_{\ttbar} = 158.1   \pm 2.1\stat \pm 10.2\syst \pm 3.5\lum\unit{pb},
\end{equation}
\end{linenomath}
which is in agreement with the QCD predictions of $164^{+10}_{-13}\unit{pb}$~\cite{Aliev:2010zk, Langenfeld:2009wd}, $163^{+11}_{-10}\unit{pb}$~\cite{Kidonakis:2010dk} and $149\pm11\unit{pb}$~\cite{Ahrens}
that are based on the full next-to-leading-order (NLO) matrix elements and the
resummation of the leading and NLO soft logarithms.

\section*{Acknowledgements}
We congratulate our colleagues in the CERN accelerator departments for the excellent performance of the LHC and thank the technical and administrative staffs at CERN and at other CMS institutes for their contributions to the success of the CMS effort. In addition, we gratefully acknowledge the computing centres and personnel of the Worldwide LHC Computing Grid for delivering so effectively the computing infrastructure essential to our analyses. Finally, we acknowledge the enduring support for the construction and operation of the LHC and the CMS detector provided by the following funding agencies: BMWF and FWF (Austria); FNRS and FWO (Belgium); CNPq, CAPES, FAPERJ, and FAPESP (Brazil); MEYS (Bulgaria); CERN; CAS, MoST, and NSFC (China); COLCIENCIAS (Colombia); MSES (Croatia); RPF (Cyprus); MoER, SF0690030s09 and ERDF (Estonia); Academy of Finland, MEC, and HIP (Finland); CEA and CNRS/IN2P3 (France); BMBF, DFG, and HGF (Germany); GSRT (Greece); OTKA and NKTH (Hungary); DAE and DST (India); IPM (Iran); SFI (Ireland); INFN (Italy); NRF and WCU (Republic of Korea); LAS (Lithuania); CINVESTAV, CONACYT, SEP, and UASLP-FAI (Mexico); MSI (New Zealand); PAEC (Pakistan); MSHE and NSC (Poland); FCT (Portugal); JINR (Armenia, Belarus, Georgia, Ukraine, Uzbekistan); MON, RosAtom, RAS and RFBR (Russia); MSTD (Serbia); SEIDI and CPAN (Spain); Swiss Funding Agencies (Switzerland); NSC (Taipei); ThEPCenter, IPST and NSTDA (Thailand); TUBITAK and TAEK (Turkey); NASU (Ukraine); STFC (United Kingdom); DOE and NSF (USA).

Individuals have received support from the Marie-Curie programme and the European Research Council (European Union); the Leventis Foundation; the A. P. Sloan Foundation; the Alexander von Humboldt Foundation; the Belgian Federal Science Policy Office; the Fonds pour la Formation \`a la Recherche dans l'Industrie et dans l'Agriculture (FRIA-Belgium); the Agentschap voor Innovatie door Wetenschap en Technologie (IWT-Belgium); the Ministry of Education, Youth and Sports (MEYS) of Czech Republic; the Council of Science and Industrial Research, India; the Compagnia di San Paolo (Torino); and the HOMING PLUS programme of Foundation for Polish Science, cofinanced from European Union, Regional Development Fund.

\bibliography{auto_generated}

\cleardoublepage \appendix\section{The CMS Collaboration \label{app:collab}}\begin{sloppypar}\hyphenpenalty=5000\widowpenalty=500\clubpenalty=5000\textbf{Yerevan Physics Institute,  Yerevan,  Armenia}\\*[0pt]
S.~Chatrchyan, V.~Khachatryan, A.M.~Sirunyan, A.~Tumasyan
\vskip\cmsinstskip
\textbf{Institut f\"{u}r Hochenergiephysik der OeAW,  Wien,  Austria}\\*[0pt]
W.~Adam, E.~Aguilo, T.~Bergauer, M.~Dragicevic, J.~Er\"{o}, C.~Fabjan\cmsAuthorMark{1}, M.~Friedl, R.~Fr\"{u}hwirth\cmsAuthorMark{1}, V.M.~Ghete, J.~Hammer, N.~H\"{o}rmann, J.~Hrubec, M.~Jeitler\cmsAuthorMark{1}, W.~Kiesenhofer, V.~Kn\"{u}nz, M.~Krammer\cmsAuthorMark{1}, I.~Kr\"{a}tschmer, D.~Liko, I.~Mikulec, M.~Pernicka$^{\textrm{\dag}}$, B.~Rahbaran, C.~Rohringer, H.~Rohringer, R.~Sch\"{o}fbeck, J.~Strauss, A.~Taurok, W.~Waltenberger, G.~Walzel, E.~Widl, C.-E.~Wulz\cmsAuthorMark{1}
% ...
\vskip\cmsinstskip
\dag:~Deceased\\
1:~~Also at Vienna University of Technology, Vienna, Austria\\
% ...

\end{sloppypar}
\end{document}